\documentclass[12pt]{iopart}
\usepackage[T1]{fontenc}
\usepackage[latin1]{inputenc}
\usepackage{graphics}
\usepackage{latexsym}
\begin{document}
\letter{Effects of the magnetic fields on the protoneutron star structure.}

\author{R. Ma\'{n}ka, M. Zastawny-Kubica, A.Brzezina, I.Bednarek}

\address{Department of Astrophysics and Cosmology,
 Institute of Physics,
 University of Silesia,
 Uniwersytecka 4, 40-007 Katowice,
 Poland.}

The equation of state of a protoneutron star matter including the
effects of magnetic field, temperature, nuclear asymmetry and
trapped neutrinos are studied. Using the
Oppenheimer-Volkoff-Tolman equation global parameters of the
protoneutron star are obtained. Presence of the nuclear asymmetry
and magnetic field make the protoneutron star more compact.

 \submitto{\JPG}

\maketitle

Properties of dense matter in strong magnetic field has been the
subject of investigations in astrophysics of white dwarfs,
protoneutron and neutron stars. It is motivated by the fact that
magnetic fields of the order of \( 10^{8} \) \( G \) - \( 10^{18}
\) \( G \) are known to exist in many cases of these objects.
Properties of protoneutron stars, under various assumptions
concerning composition and equation of state of hot lepton rich
matter were studied by many authors (Strobel et al. 1999, Bombaci
et al. 1995, Takatsuka 1995, Ma\'{n}ka et al. 2001)
\cite{Strob,Bomb,Takat,mil4}. Matter inside a protoneutron star is
highly degenerate and chemical potential of its constituents are a
few hundreds of \( MeV \).
 The strength of magnetic field of a
protoneutron star changes from \( 10^{8} \) \( G \) at the surface
up to \( 10^{18} \) \( G \) in the center of the star. The
interior magnetic fields are a few orders of magnitude larger than
the surface ones \cite{pal}. \\
 The aim of this letter is to study the basic properties of the protoneutron
stars, such as masses and radii. The construction of protoneutron
star model is based on various realistic equation of states and
results in the general picture of protoneutron star interiors.
Thus the proper form of the equation of state is essential in
determining protoneutron star properties, such as the mass-range
and the mass-radius relation. The complete and more realistic
description of a protoneutron star requires taking into
consideration effects of finite temperature and nonzero magnetic
fields. Using the relativistic mean-field theory approach the
adequate form of the equation of state enlarged by contributions
coming from magnetic field and temperature is constructed and
serves as an input to the Oppenheimer-Volkoff-Tolman equations.
The theory considered here comprises electrons, neutrinos, scalar,
vector-scalar and vector-isovector mesons. The RMF theory implies
that the nucleon interactions appear through the exchange of meson
fields \cite{ser,rei,toki,bedn}. This theory is very useful in
describing nuclear matter and finite nuclei. Its extrapolation to
large charge asymmetry is of considerable interest in nuclear
astrophysics and particulary in constructing protoneutron and
neutron star models where extreme conditions of isospin are
realized. The Lagrangian of the theory contains baryon, meson and
lepton degrees of freedom and can be represented as the sum
\begin{equation} \label{lagra}
{\mathcal{L}}={\mathcal{L}_{B}}+{\mathcal{L}_{L}}+{\mathcal{L}_{\mathcal{M}}}+{\mathcal{L}_{G}}+{\mathcal{L}_{QED}},
\end{equation}
 where \( {\mathcal{L}_{B}} \), \( {\mathcal{L}_{L}} \), \( {\mathcal{L}_{\mathcal{M}}} \),
\( {\mathcal{L}_{G}} \) describe the baryonic, leptonic, mesonic
and gravitational terms, respectively. The \( {\mathcal{L}_{QED}}
\) is the Lagrangian density function of the QED theory. The
fermion fields are composed of neutrons, protons, electrons and
neutrinos. In strong magnetic field, contributions from the
anomalous magnetic moments of protons and neutrons (\( \mu _{b}
\)) must be also considered \cite{mao}. The baryonic term of the
Lagrangian density function is given by \begin{eqnarray}
{\mathcal{L}_{B}}=i\overline{\psi }\gamma ^{\mu }D_{\mu }\psi
-\overline{\psi }(M-g_{\sigma }\varphi )\psi +\mu
_{b}\overline{\psi }\sigma _{\mu \nu }F^{\mu \nu }\psi ,
\end{eqnarray}
 where the covariant derivative \( D_{\mu } \) is defined as \begin{eqnarray}
D_{\mu }=\partial _{\mu }+\frac{1}{2}ig_{\rho }\rho _{\mu }^{a}\sigma ^{a}+ig_{\omega }\omega _{\mu }+iQA_{\mu }.
\end{eqnarray}
 The anomalous magnetic moments introduced via the minimal coupling
of the nucleons to the electromagnetic field tensor for each
baryon give contributions to the Lagrangian density function which
are contained in the \( \mu _{b}\overline{\psi }\sigma _{\mu \nu
}F^{\mu \nu }\psi  \) term , where \( \sigma _{\mu \nu
}=\frac{i}{2}\{\gamma _{\mu },\gamma _{\nu }\} \).

The leptonic part of the total Lagrangian has the form
\begin{eqnarray} {\mathcal{L}_{L}}=i\sum
_{f}\bar{L}_{f}\gamma ^{\mu }\widetilde{D}_{\mu }L_{f}-\sum
_{f}g_{f}(\bar{L}_{f}He_{Rf}+h.c),\label{lag1}
\end{eqnarray}
 where \( \widetilde{D}_{\mu }=\partial _{\mu }-ieA_{\mu } \) and
\( e \) is the electron charge. The vector potential is given by
\( A_{\mu }=\{A_{0}=0,\, A_{i}\} \) where \begin{eqnarray*}
A_{i}=-\frac{1}{2}\varepsilon _{ilm}x^{l}B_{0}^{m}.
\end{eqnarray*}
The gauge in which uniform magnetic field $B$ lies along the
z-axis was chosen $B_{0}^{m}=(0,0,B_{z})$. The mesonic part of the
Lagrangian density function consisting of isoscalar (scalar
$\sigma$, vector $\omega$) and isovector (vector $\rho$) mesons is
given by \begin{eqnarray*}
{\mathcal{L}_{\mathcal{M}}}=-\frac{1}{2}\partial _{\mu }\varphi \partial ^{\mu }\varphi -U(\varphi )-\frac{1}{4}F_{\mu \nu }F^{\mu \nu }-\frac{1}{4}W_{\mu \nu }W^{\mu \nu }-\frac{1}{2}M_{\omega }^{2}\omega _{\mu }\omega ^{\mu } &  & \\
-\frac{1}{4}c_{3}(\omega _{\mu }\omega ^{\mu })^{2}-\frac{1}{4}R_{\mu \nu }^{a}R^{a\mu \nu }-\frac{1}{2}M_{\rho }^{2}\rho _{\mu }^{a}\rho ^{a\mu }, &
\end{eqnarray*}
 where \( F_{\mu \nu } \) is
the electromagnetic stress tensor, \( W_{\mu \nu } \) and \( R^{a}_{\mu \nu } \)
are vector meson fields strength. The potential function \( U(\varphi ) \)
has a very well known form introduced by Boguta and Bodmer \cite{bod}.
The parameters used in this model are collected in Table 1.
\begin{table}
\begin{tabular}{|c|c|c|c|c|}
\hline
\( g_{\sigma }\, \,  \)&
 \( g_{\omega }\, \,  \)&
 \( g_{\rho }\, \,  \)&
 \( g_{2}[\, MeV\, ] \)&
 \( g_{3} \)\\
\hline
\( 10.0289 \)&
 \( 12.6139 \)&
\(9.2644 \)&
 \( 1427.18 \)&
 \( 0.6183 \)\\
\hline
\( M_{\sigma }\, [\, MeV\, ] \)&
 \( M_{\omega }\, [\, MeV\, ] \)&
 \( M_{\rho }\, [\, MeV\, ] \)&
 \( M\, [\, MeV\, ] \)&
 \( c_{3} \)\\
\hline
\( 511.198 \)&
 \( 783 \)&
 \( 770 \)&
 \( 938 \)&
 \( 71.3075 \) \\
\hline
\end{tabular}

\caption{\label{tab: 1}The parameter set of the model
\cite{osa7}.}
\end{table}
In the mean-field approximation baryon currents in field equations
are replaced by their ground states expectation values. Mesons
fields are replaced by their mean values \( \sigma =<\varphi > \),
\( w_{\mu }=<\omega _{\mu }>=\delta _{\mu ,0}w \) and \(
r^{a}_{\mu }=<\rho _{\mu }^{a}>=\delta ^{a,3}\delta _{\mu ,0}\, r
\). The Dirac equation for the nucleon quasiparticle has the form
\begin{equation}
(i\gamma ^{\mu }\, \overline{D}_{\mu }-m_{eff})\psi =0
\end{equation}
with the covariant derivative given by
\begin{equation}
\overline{D}_{\mu }=\partial _{\mu }+\frac{1}{2}ig_{\rho
}r^{a}_{\mu }\sigma ^{a}+ig_{\omega }w_{\mu }+iQA_{\mu }
\end{equation}
The effective nucleon mass $m_{eff}$ equals
\begin{equation} \label{mf} m_{eff}=M\delta =M-g_{s}\sigma .
\end{equation}
 This redefine the proton and neutron chemical potentials
\begin{eqnarray}
 & \mu _{p}=\varepsilon _{p}+g_{\omega }w+\frac{1}{2}g_{\rho }r & \label{mp} \\
 & \mu _{n}=\varepsilon _{n}+g_{\omega }w-\frac{1}{2}g_{\rho }r & \label{mn}
\end{eqnarray}
where \( \varepsilon_{f}=\sqrt{k^{2}+m^{2}_{eff}} \) (\( f=\{p,n\}
\)). Having obtained the effective nucleon chemical potentials one
can describe the chemical equilibrium of the system which is
imposed through relations between chemical potentials.
Protoneutron star matter is assumed to be in beta equilibrium and
electrically neutral
\begin{eqnarray}
 p+e\leftrightarrow n+\nu _{e} .
\end{eqnarray}
This can be expressed as a relation between the chemical
potentials of the protoneutron star constituents
\begin{eqnarray}
  \mu _{\nu _{e}}=\mu _{e}+\mu _{p}-\mu _{n}.
\end{eqnarray}
Using relations (\ref{mp}) and (\ref{mn}) the neutrino chemical
potential protoneutron star matter is obtained
\begin{eqnarray*}
\mu _{\nu _{e}}=\mu _{e}+\varepsilon _{p}-\varepsilon _{n}+g_{\rho
}r.
\end{eqnarray*}
The charge neutrality means that \( n_{e}=n_{p} \). The assumption
that only \( \nu _{e} \) are captured inside the star core was
made. The proton fraction is defined by the physical conditions in
the star and is determined by beta equilibrium and by nuclear
asymmetry. Effects due to muon neutrino and antineutrino were
completely ignored because muons cannot be produced in the
low-density medium. In this letter the effects of moderate
magnetic field on the equation of state of a relativistic,
degenerate gas is considered. The well known dispersion relation
for baryons takes the form \cite{Lati}
\begin{eqnarray}
E_{n,p_{z},s}^{b}=\sqrt{p_{z}^{2}+(\sqrt{m_{eff}^{2}+2nQB_{z}}+s\mu
_{b}B_{z})^{2}}\label{ener}
\end{eqnarray}
and for small values of magnetic fields strength the equation
(\ref{ener}) is given by
\begin{eqnarray*}
E_{n,p_{z}}^{b}\sim \sqrt{p_{z}^2+m_{eff}^{2}+2nQB_{z}}.
\end{eqnarray*}
 Along the field the particle motion is free and quasi-one-dimensional with the modified
density of states. The fermion density of states in the absence of
magnetic field is replaced by the sum
\begin{eqnarray*}
2\int \frac{d^{3}{p}}{(2\pi )^{3}}\rightarrow \sum_{s}\sum
_{n=0}^{\infty }[2-\delta _{n0}]\int \frac{eB_{z}}{(2\pi
)^{2}}dp_{z},
\end{eqnarray*}
 where the symbol \( \delta _{n0} \) denotes the Kronecker delta
\cite{mil14} and thus the spin degeneracy equals 1 for the ground
(\( n=0 \)) Landau level and 2 for \( n\geq 1 \). The total
pressure of the system can be described as the sum of pressure
coming from fermions, mesons plus corrections coming from
electromagnetic field
\begin{eqnarray*}
P=P_{f}+P_{{\mathcal{M}}}+P_{QED}.
\end{eqnarray*}
 The contributions from the same constituents form the energy density
\begin{eqnarray*}
\varepsilon =\varepsilon _{f}+\varepsilon
_{{\mathcal{M}}}+\varepsilon _{QED},
\end{eqnarray*}
where $\varepsilon _{QED}=P_{QED}=\frac{1}{2}B^{2}$.
 The fermion pressure is defined as
  \begin{eqnarray*}
& P_{f}=\frac{\gamma_{i} m_{i}^{4}}{4\pi ^{2}}\sum_{s}\sum
_{n=0}^{\infty }[2-\delta
_{n0}](I_{-1,2,+}(z_{i}/t_{i},(\sqrt{\delta_{i}^{2}+2\gamma_{i}n}
+s\frac{\mu_{b}}{Q_{i}}\gamma_{i})^{2}) & \\ & +
I_{-1,2,-}(z_{i}/t_{i},(\sqrt{\delta_{i}^{2}+2\gamma_{i}n}
+s\frac{\mu_{b}}{Q_{i}}\gamma_{i})^{2}) &
\end{eqnarray*}
 where the Fermi integral
 \begin{equation*}
I_{\lambda ,\eta \pm }(u,\alpha )=
\int \frac{(\alpha +x^{2})^{\lambda /2}x^{\eta }dx}{e^{\left( \sqrt{\alpha +x^{2}}\mp u\right) }+1}
\end{equation*}
 was used, \( z_{i}=\mu _{0}/m_{i} \), \( t_{i}=k_{B}T_{0}/m_{i} \),
\( u_{i}=z_{i}/t_{i} \) and \( \gamma_{i}
=B_{z}/B_{c,i}=Q_{i}B/m_{i}^{2} \). The critical magnetic field
strength for electron equals \( B_{c}=m_{e}^{2}/\mid e\mid  \)= \(
4.414\times 10^{13} \) \( G \) and for proton \( B_{c}\sim10 ^{20}
\) \( G \). The index \( i \) denotes fermions.

The fermion energy density is defined with the use of the Fermi
integral
\begin{eqnarray*}
 & \varepsilon _{f}=\frac{\gamma_{i}
m_{i}^{4}}{4\pi ^{2}}\sum_{s}\sum _{n=0}^{\infty }[2-\delta
_{n0}](I_{1,0,+}(z_{i}/t_{i},(\sqrt{\delta_{i}^{2}+2\gamma_{i}n}
+s\frac{\mu_{b}}{Q_{i}}\gamma_{i})^{2}) & \\ &
+I_{1,0,-}(z_{i}/t_{i},(\sqrt{\delta_{i}^{2}+2\gamma_{i}n}
+s\frac{\mu_{b}}{Q_{i}}\gamma_{i})^{2}). &
\end{eqnarray*}

\begin{figure}
{\par\centering \resizebox*{12cm}{!}{\includegraphics{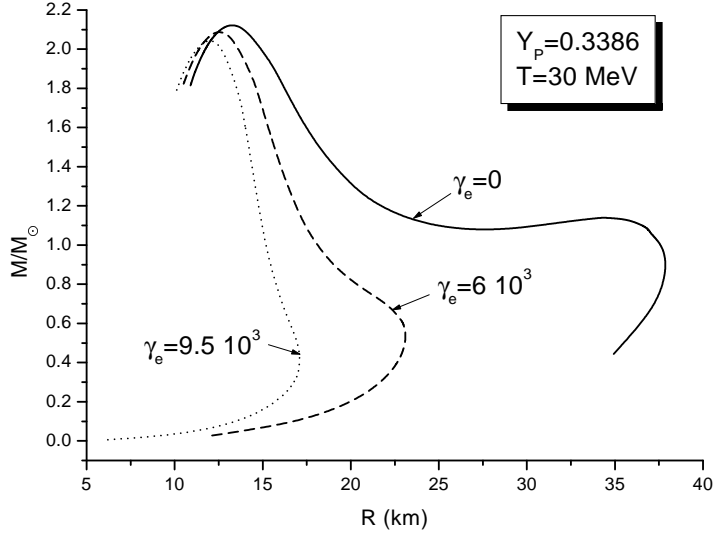}}
\par} \caption{The mass-radius relation for protoneutron star for
different values of dimensionless magnetic field strength
$\gamma_{e}=B/B_{c}$ (the critical magnetic field strength for
electron equals \( B_{c}=m_{e}^{2}/\mid e\mid  \)= \( 4.414\times
10^{13} \) \( G \)).}
\end{figure}
\begin{figure}
{\par\centering \resizebox*{12cm}{!}{\includegraphics{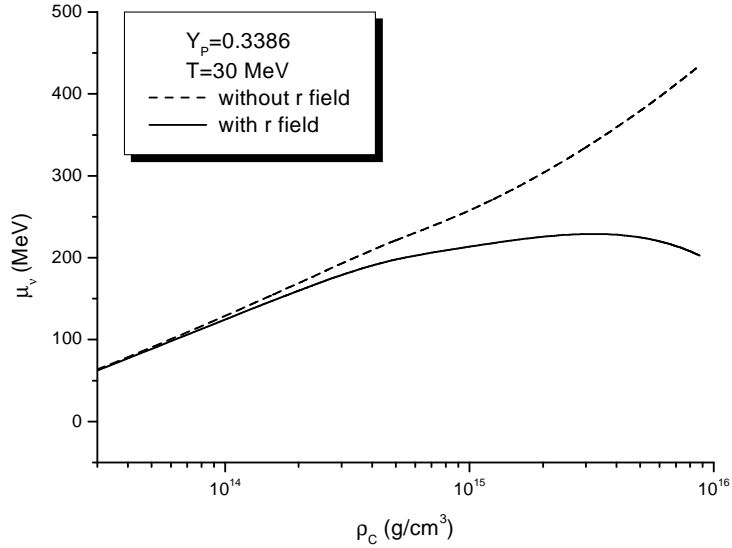}}
\par} \caption{The neutrino chemical potential versus the central density
for the protoneutron star.}
\end{figure}
\begin{figure}
{\par\centering \resizebox*{12cm}{!}{\includegraphics{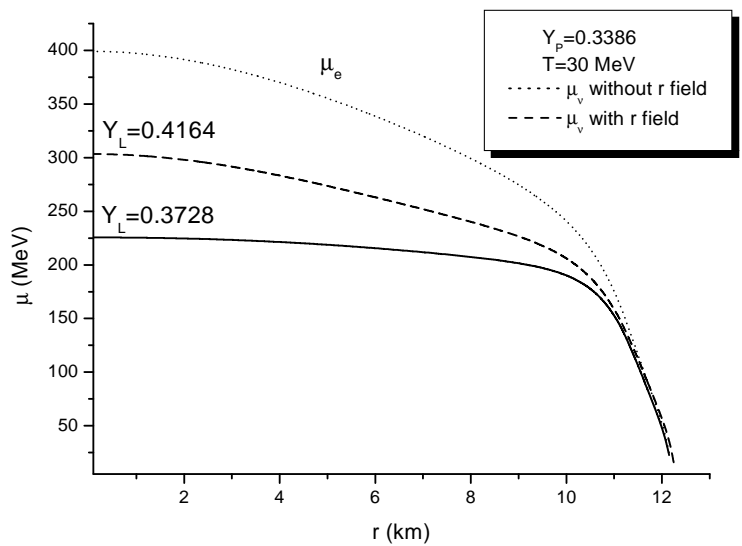}}
\par} \caption{The chemical potential-radius diagram
for the protoneutron star for different values of the lepton
fraction. The  dimensionless magnetic field strength
$\gamma_{e}=B/B_{c}=9.5 \times 10^{3}$. }
\end{figure}

The mass-radius relation for the protoneutron star with and
without magnetic field is presented in figure 1. The temperature
is fixed and equals 30 $MeV$, the lepton fraction is assumed
$Y_{L}=0.4$ and the proton fraction $Y_{P}\simeq 0.3$. These
conditions are adequate for the trapped neutrino case. In this
figure the mass-radius relations for protoneutron star with and
without field $\rho$ are depicted. The magnetic field causes that
these objects are more compact and similar to a ordinary neutron
star. Comparing protoneutron star parameters namely masses and
radii one can go to the conclusion that protoneutron star model
calculated in the presence of magnetic field results with smaller
radii and bigger masses than the one obtained in the absence of
magnetic field. The neutrino chemical potential as a function of
the central density is shown in figure 2. One can compare the
influence of the nuclear asymmetry which is driven by the meson
$\rho$. Its presence in the protoneutron star matter diminishes
the value of neutrino chemical potential and decreases the
protoneutron star mass. This feature strongly depends on the
nuclear interactions inside a protoneutron star. Figure 3 depicts
the comparision of neutrino chemical potentials calculated with
and without the influence of asymmetry with the electron one. This
picture shows the profiles of the mentioned above chemical
potentials inside the star. The nuclear asymmetry diminishes the
value of neutrino chemical potential inside the star and thus
lowers its density. The influence of asymmetry is the most
distinctive in the high density region inside a protoneutron star.
It was examined for moderate strength of magnetic field. In
conclusion, the magnetic field and nuclear asymmetry make the
protoneutron star more compact astrophisics object. \linebreak
--------------------------------------------------

\end{document}